\documentclass[]{fairmeta}
\usepackage{booktabs}
\usepackage{multirow}
\usepackage{float}
\usepackage{natbib}
\setcitestyle{numbers,square,comma}
\usepackage{wrapfig}

\usepackage{pgfplots}
\pgfplotsset{compat=1.16}
\usepgfplotslibrary{groupplots}
\usetikzlibrary{patterns}

\usepackage{tikz}
\usepackage{algpseudocode}
\usepackage[font=small,labelfont=bf]{caption}
\usepackage{array}
\usepackage{multirow}
\usepackage{subcaption}
\usepackage[normalem]{ulem}
\usepackage{xspace}
\usepackage{xparse}
\usepackage{pifont}
\usepackage[export]{adjustbox}
\usepackage[scaled=0.85]{DejaVuSansMono}
\usepackage{amssymb}

\usepackage{makecell}
\usepackage{multirow}
\usepackage{multicol}
\usepackage{subcaption}
\usepackage{colortbl}
\usepackage{color}
\usepackage{bigints}
\usepackage{textcomp}

\newcommand{\ours}{MoCA\xspace}
\newcommand{\norm}[1]{\left\lVert#1\right\rVert}

\def\dreamix{Dr$\widetilde{\text{ea}}$mix}

\title{Motion-Conditioned Image Animation for Video Editing}

\author[1,2,\dagger]{Wilson Yan}
\author[1]{Andrew Brown}
\author[2]{Pieter Abbeel}
\author[1]{Rohit Girdhar}
\author[1]{Samaneh Azadi}

\affiliation[1]{GenAI, Meta}
\affiliation[2]{UC Berkeley}

\contribution[\dagger]{Work done during an internship at Meta}

\abstract{
We introduce \ours, a \textbf{Mo}tion-\textbf{C}onditioned Image \textbf{A}nimation approach for video editing.   
It leverages a simple decomposition of the video editing problem into image editing followed by motion-conditioned image animation. Furthermore, given the lack of robust evaluation datasets for video editing, we introduce a new  benchmark that measures edit capability across a wide variety of tasks, such as object replacement, background changes, style changes, and motion edits. 
We present a comprehensive human evaluation of the latest video editing methods along with \ours, on our proposed benchmark. 
\ours establishes a new state-of-the-art, 
demonstrating greater human preference win-rate, and outperforming notable recent approaches including \dreamix{} ($63\%$), MasaCtrl ($75\%$), and Tune-A-Video ($72\%$), with especially significant improvements for motion edits.
}

\date{\today}
\correspondence{Wilson Yan at \email{wilson1.yan@berkeley.edu}}

\metadata[Website]{\url{facebookresearch.github.io/MoCA}}

\begin{document}

\maketitle

\begin{figure}[]
\centering
\includegraphics[width=\linewidth]{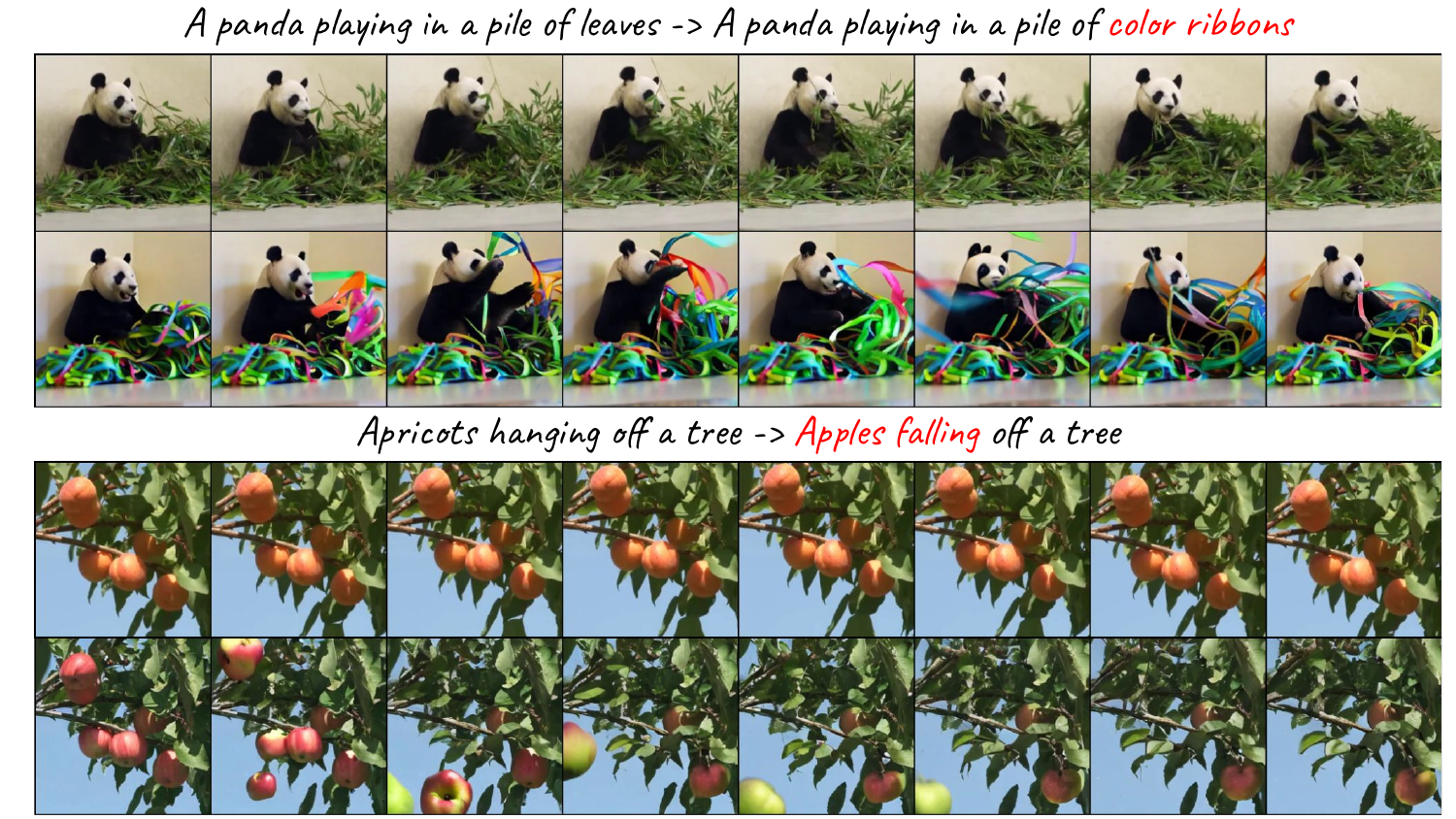}
\caption{\textbf{is able to generate a diverse range of edits, such as object replacement, style changes, and motion edits.} The frames in the top row in each example represent the source video while the bottom ones show the edited frames by \ours. The source and editing prompts are shown above each example.}
\label{fig:teaser}
\end{figure}

\section{Introduction}
\label{sec:intro}
Recent advancements in image and video generation models have seen tremendous progress, with existing models able to synthesize highly complex images~\cite{dalle,dalle2,SD,imagen,muse} or videos~\cite{phenaki,make-a-video,alignyourlatents,imagenvideo,emuvideo} given textual descriptions. 
Outside of generating purely novel content, these models have shown to be powerful tools in achieving advanced image and video editing capabilities for downstream content creation. 

Given a source video, a caption of the source video, and an editing textual prompt, a video editing method should produce a new video that is aligned with the provided editing prompt while retaining faithfulness to all other non-edited characteristics of the original source video. Video edit types can be broadly split into two main categories of spatial and temporal edits. Spatial edits generally consist of image-based edits extended to video, such as editing a video in the style of Van Gogh, inserting an object into the scene, or changing the background. Due to the added temporal dimension in video, we can also change the underlying motion of the object, such as making a panda play in a pile of ribbons, or replacing apricots in a video with apples and making them fall off a tree (see \Cref{fig:teaser}).

Current methods in video editing focus more on spatial editing problems while ignoring the motion editing problem. Proposed methods leverage pre-trained text-to-image or video models for editing by further fine-tuning with conditioning on auxiliary information such as depth maps or edge maps~\cite{gen1,controlnet}, fine-tuning for each edit example~\cite{tav,dreamix,dreambooth,imagic,textualinversion}, or exploiting the diffusion process to restrict the generated edits to share similar features and structures with the source content~\cite{p2p,videop2p,masactrl,pnp,sdedit,rerender,tokenflow}. Notably, most proposed methods either in image or video editing are generally specialized only to a subset of editing tasks and do not perform well on others. For example, methods to utilize depth or edge maps of the video~\cite{gen1,controlnet} find it more difficult to perform motion edits due to adherence to the original video structure. As such, it becomes important to assess video editing capabilities across a wide range of different edits in order to better understand their advantages and disadvantages.

In this paper, we present two key contributions for video editing. 
First, we introduce a simple yet strong approach 
for video editing, \textbf{Mo}tion-\textbf{C}onditioned Image \textbf{A}nimation (\ours), that decomposes the problem into image editing and image animation. We first use the existing image editing methods to edit the first video frame, then produce an edited video using a motion-conditioned image animation model. We use an optical flow representation of the source video using a pretrained RAFT~\cite{raft} model as the motion conditioning to retain the original motion characteristics of the source video. In the video edits consisting of a motion edit, we drop out this motion conditioning.
Through our extensive experiments and human evaluations, we show that this simple baseline outperforms the state-of-the-art video editing models across a wide range of edit types. 

Secondly, we introduce a dataset of 250+ video edits that comprehensively covers a wide range of video editing types. We combine existing datasets for video editing, and introduce our own subset of curated videos from YouTube-8M~\cite{youtube8m} with a stronger emphasis in including motion-based edits due to a general lacking of such examples in current public video editing datasets~\cite{loveu}. 
Using our combined dataset, we comprehensively benchmark prior video editing methods along a range of pre-categorized edits types, such as style, background, object, and motion-based edits via human evaluation and automatic metrics. Additionally, we perform an analysis of the alignment between the automatic metrics for measuring video editing quality and human judgement.

\section{Related Work}
\label{sec:related_work}
By the remarkable progress in text conditional image and video generation models, text-guided image and video editing have emerged as key editing tools that enable average users and artists to create new content easily from existing photos or videos. 
In this section, we will discuss the existing works on diffusion-based text-driven image and video editing and their applicability to various manipulation tasks.

\subsection{Text-driven Image Editing}

Prior works have proposed a variety of methods for text-conditioned image editing. One family of such image editing methods focus on using diffusion models, and produce image edits by altering the backward diffusion process. SDEdit~\cite{sdedit} is a simple image editing approach that applies various diffusion noise levels to an input source image, and produces image edits by sampling back out through the diffusion process conditioned on an edit prompt.
Plug-and-Play~\cite{pnp} samples edited videos initialized from the DDIM~\cite{ddim} inversion, with selected visual features copied between the source and generated images during diffusion sampling. 
Prompt-to-Prompt~\cite{p2p} (P2P) enables general edit types (style changes, object replacement, changing texture) through replacing self and cross attention of the generated image with the attention maps of the source image during diffusion sampling. 
Null-Text Inversion~\cite{nti} extends P2P to enable better editing performance when editing real images through optimizing null text embeddings to allow for more faithful reconstruction of the source image during diffusion sample.

Another class of image editing techniques that allow for more global changes in visual features are built on ControlNet~\cite{controlnet} or T2I-Adapters~\cite{t2iadapter}, where pretrained text-to-image models are augmented and finetuned to incorporate conditioning information, such as depth maps or contour maps computed from edge detection algorithms.

Most prior image editing methods are generally constrained structurally, and have a more difficult time producing image edits with large pose changes, such as editing an image of a bird to spread its wings. MasaCtrl~\cite{masactrl} achieves this through mutual self-attention, where select self-attention layers in the diffusion networks attend to the keys and values of the corresponding layers of the source image during the diffusion process. Larger pose changes are enabled by only enabling mutual self-attention replacement during later diffusion timesteps. Imagic~\cite{imagic} similarly achieves image edits with larger pose changes through a text embedding optimization and model fine-tuning process.

Lastly, the Instruct-Pix2Pix~\cite{instructpix2pix} family of models propose to treat the image editing problem as a supervised learning problem. Core work around these model requires collecting supervised data as pairs of (text editing instructions and images before/after
the edit), and fine-tuning a pre-trained text-to-image model on the collected data.

In this paper, we focus on video editing, as it provides a more challenging task in accomplishing complex edits over both space and time.

\subsection{Text-driven Video Editing}
Video editing similarly can be deployed for various manipulation tasks including style transfer, object or scene manipulations, and motion editing. However, these manipulation tasks are more challenging in videos since the generated content should be consistent across frames.
Most of the existing works in video editing focus more on the first two manipulation types while ignoring the motion editing problem, as they generally propose video editing methods that leverage pretrained text-to-image models. 
Pix2Video~\cite{pix2video} and FateZero~\cite{fatezero} both propose different variants of extending self-attention with cross-frame attention.
TokenFlow~\cite{tokenflow} Rerender a Video~\cite{rerender}, and CoDef~\cite{codef} perform video edits through image editing techniques and propagating edited features temporally using estimated motions by computing temporal inter-frame correspondences~\cite{tokenflow}, optical flow estimation and warping~\cite{rerender}, or estimating canonical images and temporal deformation fields using optical flow of the source video~\cite{codef}, respectively.

Methods such as Gen-1~\cite{gen1}, VideoComposer~\cite{videocomposer}, and ControlVideo~\cite{controlvideo} train text-to-video models with additional conditioning inputs such as depth maps or motion vectors to allow for controllability of general structure in the resulting video edits.

Tune-a-Video~\cite{tav} and Dreamix~\cite{dreamix} both propose fine-tuning a pre-trained diffusion model for each source video, where Tune-a-Video finetunes a pretrained text-to-image model and Dreamix fine-tunes a pre-trained text--to-video model. For both methods, edits are produced using the fine-tuned video model to sample back out conditioned on given the edit prompts.

Most prior works tend to target specific types of edits and evaluate on their own constructed sets of edit prompts. As such, the benefits of each method across different kinds of edits are less clear, and motivates us to propose a benchmark centered around a more rigorous analysis of the pros and cons of each method.
In addition, we propose our own video editing method that leverages existing text conditional image editing models to edit the first frame of a source video and extrapolate its future frames via a motion conditional video generation diffusion model to enable alignment of the edit with the source video and the editing prompt.

\section{Background}
\label{sec:background}
\par \noindent \textbf{Conditional latent diffusion models.} Diffusion models learn to generate samples from a training distribution by reversing a gradual noising process. At the sampling time starting from a Gaussian noise, the model generates less noisy samples in T time-steps where each time-step, $t$, corresponds with a specific noise level~\cite{dhariwal2021diffusion}. 
In latent diffusion models for each input image $x$, this noising and denoising process is applied on the latent space, $z = \mathcal{E}(x) $, of a pretrained variational autoencoder with encoder $\mathcal{E}$, resulting in more efficient training and sampling steps. In a text-conditional latent diffusion model, text features are extracted from a pre-trained language model, and then fed into the latent diffusion U-Net blocks via cross-attention modules. In addition to the text conditioning, the image or video generation models can be also conditioned on an additional input image by concatenating its features with the noisy latent features at each time-step, $z_t$, and adding extra input channels to the first convolutional layer of the U-Net~\cite{instructpix2pix,emuvideo,zeng2023make}. The network will be trained to predict the noise added to the noisy latent features given image and text conditioning inputs, respectively.

\par \noindent \textbf{Classifier free guidance.} 
Classifier-free guidance was proposed in~\cite{cfg} and is widely used to improve the fidelity and diversity of the generated samples and their correspondence with the conditioning input in a diffusion model. During training, the diffusion model is trained jointly in a conditional and unconditional setting where the conditioning input is set to NULL with a specific frequency. At inference, the generated samples are guided to be more faithful to the conditioning input while being further away from the NULL input with a guidance scale $s >= 1$. 

\begin{figure*}
    \centering
    \includegraphics[width=\textwidth]{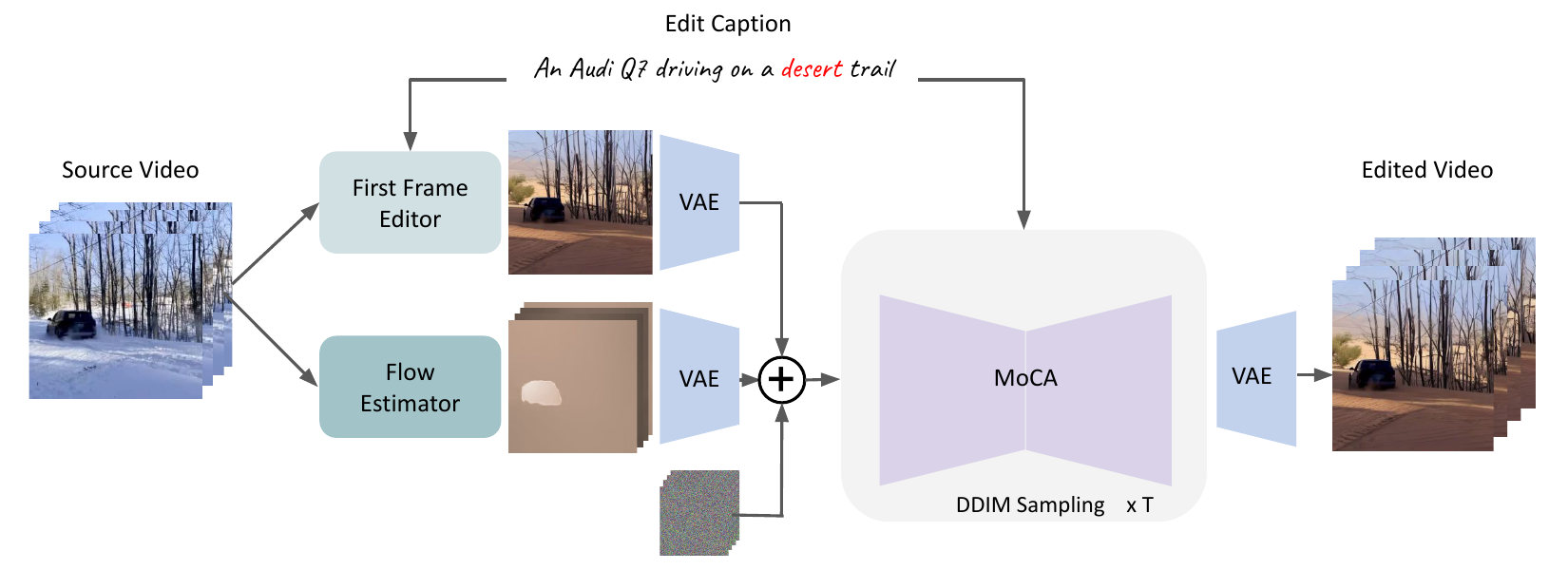}
    \caption{\textbf{An overview of \ours.} Given a source video, we compute its optical flow, and apply image editing techniques on the first frame. To produce the resulting video edit, we sample our model conditioned on motion, the edited first frame, and the edit caption. For motion-based edits, we dropout the optical flow conditioning.}
    \label{fig:method}
\end{figure*}

\section{\ours}
\label{sec:method}
Inspired by the success of image conditioning for video generation~\cite{emuvideo,zeng2023make} and text-driven image editing methods~\cite{p2p,ledits,instructpix2pix,dreambooth,controlnet}, we introduce a simple yet strong baseline for text-driven video editing that can be deployed in a wide range of video editing applications. We decompose the video editing problem into image editing, and motion-conditioned image animation. We choose to adopt this decomposition, as (1) image editing has shown large success, with the growing availability of more capable image editors that are able to edit a variety of complex editing prompts, and (2) the recent success of image animation methods for video generation to model temporal dynamics in videos~\cite{emuvideo,zeng2023make}. As a simple approach, we could first leverage image editing techniques to edit the first frame given the editing prompt, and then use an image animation model to predict the rest of the frames. However, this typically results in videos that diverge from the motion of the original source video, which is important to retain especially for edits that only target to change the style, or background of the video. Therefore, for these cases, we propose to additionally condition on a motion representation of the video. When editing video motion, we dropout conditioning on the motion and predict future frames conditioned only on the edited image. An overview of \ours is presented in \Cref{fig:method}.

To achieve this, we train a latent diffusion video generation model conditioned on (1) a text prompt, (2) an image as the first frame of each video, and (3) an optical flow representing the motion in the video. We use a variational autoencoder (VAE) pre-trained on an internal image database to encode an input video with shape $T \times 3 \times H \times W$ frame-wise to a tensor of shape $T \times C \times H' \times W'$. We learn the diffusion process on this latent space and finally decode the latent to the video pixel space via its decoder. We use a pre-trained Flan-T5-XXL~\cite{flan-t5} text encoder to extract text features, $c_T$, and feed them into the model via cross-attention modules. At inference time, we edit the first frame of the source video via off-the-shelf text-based image editing models~\cite{p2p,sdedit} and use that as the image conditioning input while using the optical flow estimation of the source video as the motion conditioning. In the case where motion edits are desired, we dropout the motion-conditioning.

\paragraph{Image Editing}
We leverage a diverse range of existing image editing methods, and find certain editing methods to be useful for specific editing types. We have found Prompt-to-Prompt~\cite{p2p} effective for style, background, and object replacement edits, and SDEdit~\cite{controlnet} for multi-spatial edits that contain larger feature and pose changes. For motion-only edits, we keep the original source frame. 

\par \noindent \textbf{Image conditioning.} Inspired by Emu-video~\cite{emuvideo}, to condition the video generation model on the first frame as an input, $c_I$, we encode the first frame using the same auto-encoder, $\mathcal{E}(c_I)$, repeat it $T$ times to have the same shape as the video latent features, and then concatenate it channel-wise with the noisy latent features at each time-step.

\par \noindent \textbf{Motion conditioning.} To condition the video generation model on the optical flow motion representation, $c_M$, we convert the optical flow into an RGB video, encode each frame using the same auto-encoder, $\mathcal{E}(\text{toRGB}(c_M))$ and concatenate it channel-wise with the noisy latent and image conditioning features at each time-step. 
Since conversion to RGB re-normalizes the frames, we additionally compute an average flow magnitude term, and condition the video model on it. This conditioning is performed similar to the diffusion time-step conditioning.

\par \noindent \textbf{Classifier free guidance for three conditionings.} Similar to~\cite{instructpix2pix,emuvideo}, we leverage classifier-free guidance with respect to all three conditioning inputs to control faithfulness to each of the inputs at inference time. During training, we randomly set each of the individual conditioning inputs and their pairwise combination to NULL for 10\% of the examples, respectively. To train for both motion-conditioned image animation, and image animation only, we train with 50\% dropout on motion conditioning. We have therefore, three guidance scales for the text, image, and motion conditioning inputs that we adjust based on the editing application. During inference, we use the following conditioning order to compute the classifier guidance, where $v_\theta$ is the output from the U-Net, and $\tilde{v}_\theta$ is used to denoise the input image:
\begin{align*}
    \tilde{v}_\theta(z_t, c_M, c_T, c_I) &= v_\theta(z_t, \varnothing, \varnothing, \varnothing) \\
    &+ s_I \cdot (v_\theta(z_t, \varnothing, \varnothing, c_I) - v_\theta(z_t, \varnothing, \varnothing, \varnothing)) \\ 
    &+ s_T \cdot (v_\theta(z_t, \varnothing, c_T, c_I) - v_\theta(z_t, \varnothing, \varnothing, c_I)) \\ 
    &+ s_M \cdot (v_\theta(z_t, c_M, c_T, c_I) - v_\theta(z_t, \varnothing, c_T, c_I)) \\ 
\end{align*}

\section{Experiments}
\label{sec:experiments}

\subsection{Implementation Details}

Our video generation model is built from a text-to-image U-Net based latent diffusion model pre-trained on our internal database of 400M (image, text) pairs. Similar to earlier works in video generation~\cite{make-a-video,alignyourlatents,emuvideo}, we expand this 2D U-Net to video generation by adding temporal modules consisting of 1D temporal convolution layers and 1D temporal attention blocks after each spatial convolution and attention block, respectively. 

We initialize all the spatial parameters from the pre-trained text-to-image model and fine-tune all the temporal and spatial layers of our video prediction model, with 1.4B trainable parameters, on an internal licensed dataset consisting of 34M pairs of video-text samples. We sample random $256 \times 256$ 2-second clips from each video using a frame rate of 4 frames per second with the first frame as the conditioning image. Videos are encoded via the pre-trained VAE to the $4\times 8 \times 32\times 32$ resolution. We additionally use RAFT~\cite{raft} to extract an estimated optical flow representation for each video during training. Similar to ~\cite{emuvideo}, we train our model using zero terminal-SNR and v-prediction, on a batch size of 512 split across 32 A100 GPUs. During inference we use 64 DDIM steps for sampling.  

\subsection{Evaluation Dataset}
\label{sec:data}
\begin{wraptable}{r}{0.5\textwidth}
\vspace{-1cm}
\begin{tabular}{@{}lccc@{}}
\toprule
\multicolumn{1}{c}{}         & \textbf{\begin{tabular}[c]{@{}c@{}}LOVEU\\ TGVE\end{tabular}} & \textbf{\begin{tabular}[c]{@{}c@{}}Dreamix \\ Dataset\end{tabular}} & \textbf{\begin{tabular}[c]{@{}c@{}}Our \\ Dataset\end{tabular}} \\ \midrule
\textbf{Style}               & 35                                                            & 1                                                                   & 11                                                              \\
\textbf{Background}          & 35                                                            & 1                                                                   & 7                                                               \\
\textbf{Object}              & 35                                                            & 7                                                                   & 14                                                              \\
\textbf{Motion}              & 0                                                             & 2                                                                   & 68                                                              \\
\textbf{Multi-Spatial}       & 35                                                            & 0                                                                   & 0                                                               \\
\textbf{Multi-Motion}        & 0                                                             & 0                                                                   & 17                                                              \\
\textbf{Total}               & 140                                                           & 14                                                                  & 117                                                             \\ \midrule
\textbf{\# Unique Videos}    & 35                                                            & 9                                                                   & 37                                                              \\
\textbf{Avg Edits Per Video} & 4                                                             & 1.56                                                                & 3.16                                                            \\ \bottomrule
\end{tabular}
\caption{\textbf{Details for each individual dataset in the VideoEdit benchmark, as well the total distribution of edits types.} Our custom curated dataset focuses more heavily on motion editing due to the general lack of motion edits available in the combined LOVU-TGVE and Dreamix datasets.}
\label{table:dataset_info}
\vspace{-1cm}
\end{wraptable}
We introduce a dataset of 271 edit tasks, defined as a set of (source video, edit prompt) pairs designed to comprehensively evaluate and benchmark video editing capabilities of current methods. 
Our dataset consists of a combination of existing video editing datasets, as well as our own curated subset:
\begin{itemize}
    \item \textbf{LOVEU-TGVE Dataset}~\cite{loveu}: comprises of 35 source videos, with 4 different manipulation tasks proposed for each video (140 edits total). We filtered out videos with human faces and hands.
    \item \textbf{Dreamix Dataset}~\cite{dreamix}: consists of 14 videos downloaded from the dreamix paper website, with edits primarily focusing on scene changes with motion. 
    \item \textbf{Our Custom Dataset}: we curate an additional 37 videos from YouTube-8m~\cite{youtube8m}, focused on including a diverse range of motion edits as well as a composition of scene and motion edits (117 edits total). 
\end{itemize}
We group each edit task into one of following edit types, some of which are explored in \cite{loveu,dreamix}:
\begin{itemize}
    \item \emph{Style}: changes in the composition of the video, such as making the video reflect a specific artistic style (crayon drawing, oil painting, impressionism),
    \item \emph{Object}: adding or replacing objects in the scene, such as replacing a lion with a zebra, or placing a hat on a person's head,
    \item \emph{Background}: changes in the background scene of the video, such as replacing a snowy mountain background with a desert,
    \item \emph{Motion}: changes in the motion of an entity compared to the source video, such as making a monkey jump, or a car turn in a different direction,
    \item \emph{Multi-Spatial}: a combination of style, object, and background changes in the video,
    \item \emph{Multi-Motion}: a combination of style, object, and background changes in addition to a motion change.
\end{itemize}
\Cref{table:dataset_info} shows a break-down of the number of videos and edits of each type for each dataset.

\begin{figure}
    \centering
    \includegraphics[width=\textwidth]{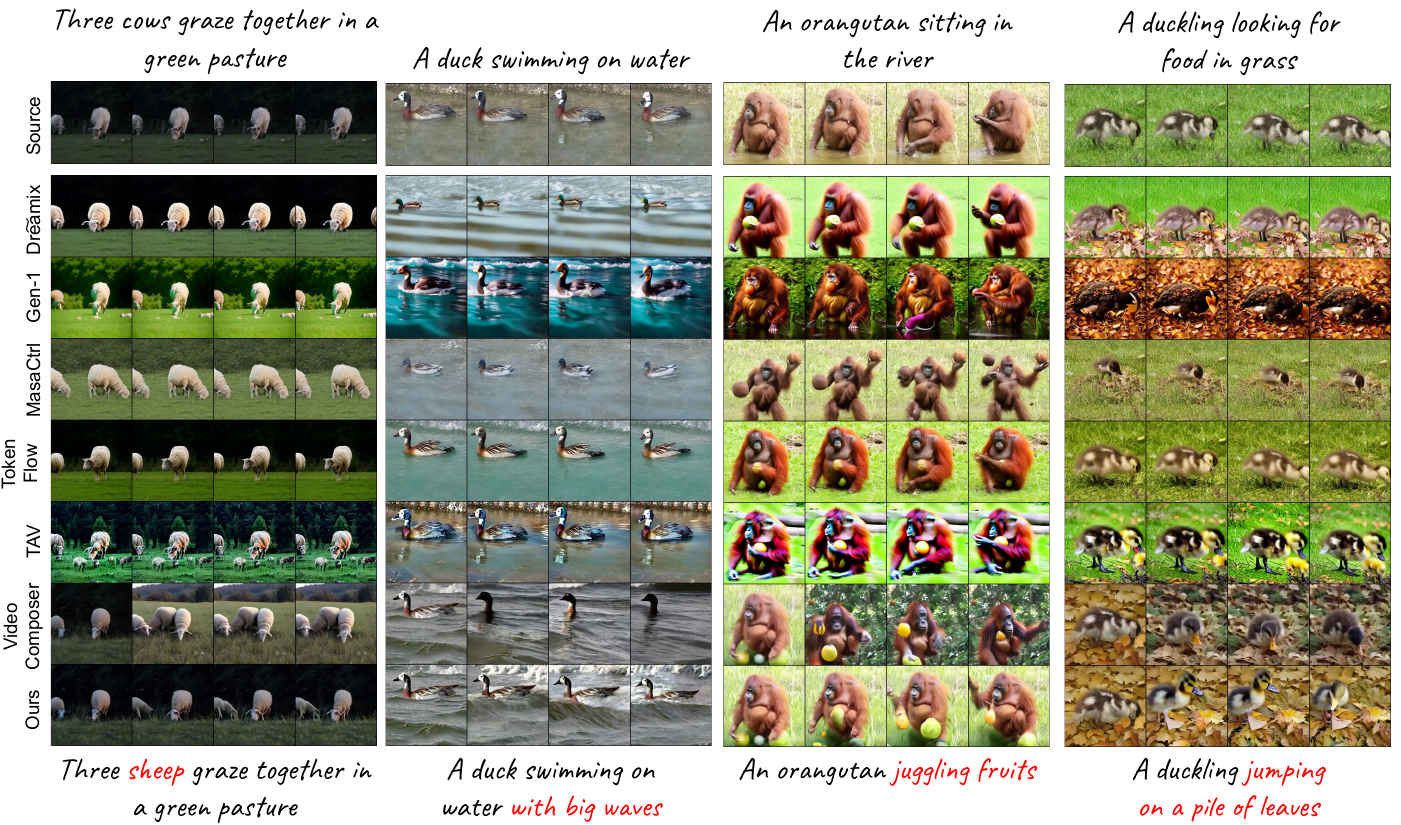}
    \caption{\textbf{Comparison of our method against baselines for a given video editing task.} Our method is able to accurately edit both the spatial and temporal properties of the source video.}
    \label{fig:method_comparison}
\end{figure}

\subsection{Baselines}
We compare against a set of SOTA baselines to comprehensively target different families of video editing models.
\begin{itemize}
    \item \textbf{TokenFlow} is a tuning-free text-to-image based editing model~\cite{tokenflow}. It leverages a pre-trained text-to-image diffusion model to edit videos by computing and propagating spatial edits across temporal correspondences found in the original source videos. We use the public repo~\footnote{\url{https://github.com/omerbt/TokenFlow}} to run this baseline. 
    \item \textbf{Tune-a-Video} is a fine-tuning text-to-image based editing model~\cite{tav}. For each edit task, Tune-a-Video extends a pre-trained text-to-image diffusion model to the temporal domain and fine-tunes the weights on a given source video. During inference, the edit result is generated through a DDIM initialized sampling using the fine-tuned model conditioned on the edit prompt. We use the public repo~\footnote{\url{https://github.com/showlab/Tune-A-Video}} to run this baseline.
    \item \textbf{\dreamix} is a fine-tuning text-to-video based editing model~\cite{dreamix}. It fine-tunes a text-to-video model for \emph{each source video}, and edits are generated by sampling at different levels of noise strengths, conditioned on the edit prompt. Due to the lack of public code and models, we perform Dreamix fine-tuning using our own internal text-to-video model (as indicated by the tilde). Our text-to-video model follows the same training parameters and model architecture (LDM) as \ours, only without the initial concatenating for image and motion conditioning.
    \item \textbf{MasaCtrl} is a tuning-free text-to-video based editing model~\cite{masactrl}. Originally presented in the text-to-image domain, MasaCtrl enables more robust structural (e.g. pose) changes in image edits compared to the prior methods. It replaces self-attention layers with mutual self-attention to query correlated local structures and textures from a source image. We extend MasaCtrl to the video domain using our text-to-video model (same as that of the {\dreamix} baseline). Importantly here, we found it crucial to only apply mutual self-attention layers in the spatial, and not temporal transformer layers of our network.
    \item \textbf{Gen-1} is a tuning-free video editing model~\cite{gen1}. Gen-1 video-to-video model generates a video conditioned on a given edit prompt and a depth map of the source video. We use the public web interface in generating edit results. 

    \item \textbf{VideoComposer} is a tuning-free text-to-video generation model~\cite{videocomposer}. VideoComposer is a motion-conditioned video model that can generate videos conditioned on a single image, and desired motion extracted from the source video. We use the public repo~\footnote{\url{https://github.com/damo-vilab/videocomposer}} to run this baseline. When generating edits, we condition VideoComposer on the same edited image as given to our method.

\end{itemize}
All baselines are run in their native resolutions and frame rates, and spatio-temporally down-sampled to 256 resolution, 4 frames per second for fair comparisons to our method. For each method and (video, edit prompt) pair, we perform a hyper-parameter sweep to generate 10 candidate edits, and use human evaluators to select the best edit. The hyper-parameter sweeps vary for each method, with some over one to three hyper-parameters while still restricted to the same max 10 candidate generations.

\subsection{Human Evaluation} 

\paragraph{Methodology}
We use crowd-sourced workers from Amazon Mechanical Turk (AMT) for our human evaluations. For each task given the source video and the editing prompt, evaluators are asked to perform a binary selection on their preferred video edit out of two given edits, one from our proposed method. 
Inspired by the JUICE metric introduced in~\cite{emuvideo}, they are also required to choose the reasoning for their selection as either a better consistency with the source video or a higher alignment with the editing prompt or both. 
The same task is given to five different evaluators, and the overall preferred video edit is selected through a majority vote. 
We evaluate paired comparisons between our method and all given baselines, and report the final metrics as the percentage of video edit examples for which our method is preferred.

\begin{table}[]
\centering
\begin{tabular}{@{}lcccccccc@{}}
\toprule
\multicolumn{1}{c}{}  & \textbf{Style} & \textbf{Background} & \textbf{Object} & \multicolumn{1}{l}{\textbf{Motion}} & \multicolumn{1}{l}{\textbf{Multi-Spatial}} & \multicolumn{1}{l}{\textbf{Multi-Motion}} & \multicolumn{1}{l}{\textbf{Total}} \\ \midrule
\textbf{\dreamix~\cite{dreamix}}       & 53\%           & 53\%                & 63\%            & 81\%                                & 49\%                                       & 65\%                                      & 63\%                               \\
\textbf{Gen-1~\cite{gen1}}         & 66\%           & 40\%                & 80\%            & 99\%                                & 57\%                                       & 90\%                                      & 74\%                               \\
\textbf{MasaCtrl~\cite{masactrl}}      & 74\%           & 72\%                & 80\%            & 76\%                                & 71\%                                       & 65\%                                      & 75\%                               \\
\textbf{Tune-a-Video~\cite{tav}}  & 53\%           & 67\%                & 70\%            & 86\%                                & 74\%                                       & 85\%                                      & 72\%                               \\
\textbf{TokenFlow~\cite{tokenflow}}     & 70\%           & 77\%                & 63\%            & 83\%                                & 83\%                                       & 85\%                                      & 76\%                               \\
\textbf{VideoComposer~\cite{videocomposer}} & 78\%           & 76\%                & 92\%            & 84\%                                & 74\%                                       & 85\%                                      & 82\%           \\ \bottomrule                   
\end{tabular}
\caption{\textbf{Human evaluation results for preference of our method over each of the baselines.} User ratings generally show greater preference for our method, with the exception of Gen-1 for background edits, and {\dreamix}  for multi-spatial edits.
}
\label{table:main_by_style}
\end{table}

\paragraph{Results}
\Cref{table:main_by_style} shows human evaluation results comparing our method against each of the baselines, partitioned by edit type. A value of 50\% means that both methods perform equally, with values greater than 50\% showing a stronger human preference towards the edits produced by our method. Evaluators significantly preferred our method over all other baselines. When examining results split by edit type, human raters showed a larger gap in preference for our method's motion edits, with a more narrow gap on spatial edits. Gen-1 notably shows capabilities in background edits, as seen by the 40\% preference for our method, and 60\% for Gen-1.
We hypothesize that since background edits require larger visual feature deviations from the original source video, Gen-1 performs well on this task by generating high quality videos. However, it is less preferred on other video edit types since it does not preserve the visual features and style of the source video due to only conditioning on depth maps. 

{\dreamix} shows similarly competitive results in the spatial edits, but struggles more on the motion edits. We found that {\dreamix} has a tendency to overfit to the motion of the source video, even when adjusting the number of fine-tuning steps. MasaCtrl shows strong motion-editing abilities, but struggles more with spatial edits, especially those with larger feature changes, due to its reliance on mutual self-attention on visual features of the source content. Tune-a-Video has strong spatial editing capabilities, but generates less temporally coherent motion, and performs poorly on the motion edits. Similarly, TokenFlow has a reasonable performance in spatial edits, but struggles with motion edits due to its reliance on a pre-trained text-to-image model. Lastly, VideoComposer generally struggles to remain faithful to the image conditioning input, or produces less temporally coherent motions.

\begin{table}[]
\centering
\begin{tabular}{@{}clccccccc@{}}
\toprule
\multicolumn{1}{l}{}        & \multicolumn{1}{c}{} & \textbf{Style} & \textbf{Background} & \textbf{Object} & \multicolumn{1}{l}{\textbf{Motion}} & \multicolumn{1}{l}{\textbf{Multi-Spatial}} & \multicolumn{1}{l}{\textbf{Multi-Motion}} & \multicolumn{1}{l}{\textbf{Total}} \\ \midrule
\multirow{3}{*}{ImageCLIP} & $\text{M}_\text{sim}$            & 45\%           & 43\%                & 47\%            & 63\%                                & 44\%                                       & 50\%                                      & 51\%                               \\
                            & $\text{M}_\text{dir}$      & 80\%           & 72\%                & 74\%            & 53\%                                & 81\%                                       & 66\%                                      & 68\%                               \\
                            & $\text{M}_\text{geo}$      & 78\%           & 71\%                & 70\%            & 50\%                                & 82\%                                       & 67\%                                      & 67\%                               \\ \midrule
\multirow{3}{*}{VideoCLIP} & $\text{M}_\text{sim}$            & 42\%           & 44\%                & 53\%            & 74\%                                & 40\%                                       & 51\%                                      & 55\%                               \\
                            & $\text{M}_\text{dir}$      & 78\%           & 74\%                & 77\%            & 59\%                                & 76\%                                       & 73\%                                      & 72\%                               \\
                            & $\text{M}_\text{geo}$      & 79\%           & 72\%                & 77\%            & 56\%                                & 71\%                                       & 78\%                                      & 69\%                               \\ \bottomrule
\end{tabular}
\caption{\textbf{Classification accuracy of each CLIP-based automatic metric}, considering binary human decisions comparing \ours edits against different baselines as the ground truth labels. Note that random guessing achieves roughly 50\% accuracy. $\text{M}_\text{dir}$ and $\text{M}_\text{geo}$, standing for CLIP text-video directional and geometric similarity scores, show relatively high accuracy (up to 80\%) on spatial-based edits, such as style, background, object, and multi-spatial. However, both methods have a much harder time selecting the correct motion-based edits.}
\label{table:metric_classification}
\end{table}

Lastly, \Cref{fig:human_factor_ratings} shows the distribution of factors selected in which human raters preferred our method over baselines. In general human raters preferred MoCA due to its stronger alignment with the edit prompt. VideoComposer shows a slightly different distribution, of which we hypothesize may be due to the fact that it would generally produce videos with high text-video alignment, but may deviate far from the source or edited image, thus the higher distribution in selecting consistency with the source video as a deciding factor.
An example video edit by all models is shown in \Cref{fig:method_comparison}.

\subsection{Automatic Evaluation}
\label{sec:automatic_evals}

In addition to presenting human evaluation results, we also investigate automatic evaluation metrics using pre-trained Video and Image CLIP models~\cite{clip,internvideo}. We perform an analysis over several CLIP-based metrics to measure video editing quality. 

\par \noindent \textbf{CLIP video similarity score.} Given a source video $\text{V}_{\text{source}}$ and an edit result $\text{V}_{\text{edit}}$, we first measure faithfulness between the source video and the resulting edited video:
\begin{align*}
    \text{M}_\text{sim} = \mathcal{E}_V(V_{\text{source}}) \cdot \mathcal{E}_V(V_{\text{edit}})
\end{align*}
where $\mathcal{E}_V$ is the VideoCLIP encoder. 
\par \noindent \textbf{CLIP text-video directional similarity score.} Given a source prompt $\text{T}_{\text{source}}$ and an edit prompt $\text{T}_{\text{edit}}$, we measure the edit quality using a CLIP text-video directional similarity metric~\cite{stylegannada,instructpix2pix}, defined as:
\begin{align*}
    &\Delta T = \mathcal{E}_T(\text{T}_{\text{edit}}) - \mathcal{E}_T(\text{T}_{\text{source}}) \\
    &\Delta V = \mathcal{E}_V(\text{V}_{\text{edit}}) - \mathcal{E}_V(\text{V}_{\text{source}}) \\
    &\text{M}_\text{dir} = \frac{\Delta V \cdot \Delta T}{\norm{\Delta V}_2\norm{\Delta T}_2}
\end{align*}
where $\mathcal{E}_T$ is the VideoCLIP text encoder. This metric measures the consistency of the change between the two videos (in CLIP space) with the change between the two prompts.

\par \noindent \textbf{CLIP text-video geometric similarity score.} Lastly, since both metrics are important in measuring the overall editing quality~\cite{instructpix2pix}, we consider an additional metric consisting of the geometric average of both metrics which would be penalized if one of the metrics is too low.
\begin{align*}
    &\text{M}_\text{geo} = \sqrt{\text{M}_\text{sim} * \text{M}_\text{dir}}
\end{align*}

We similarly compute these scores using a pre-trained image-CLIP encoder as the average of per-frame similarity scores.

\par \noindent \textbf{Correlation between automatic and human scores.} In order to measure the alignment of each evaluation metric to the human judgements, we treat the paired edit selection task as a binary classification problem, where for each pair of given video edits, we compute the ground-truth label as the majority vote among human raters. \Cref{table:metric_classification} shows the classification results for each of the automatic evaluation metrics using both Image and Video-based CLIP models.We use the original L/14 Image CLIP model, and a VideoCLIP model introduced in \cite{internvideo}. 
Note that random guessing would achieve roughly 50\% accuracy. 

Both encoder models show similar trends across all metrics, where even the highest measured overall accuracy (72\%) for $\text{M}_\text{dir}$ using VideoCLIP is still far off from perfectly aligning with human judgement. 
When split by edit type, metrics computed for spatial-type edits (style, background, object, multi-spatial) are most aligned with human raters (up to 80\% accurate), whereas motion-based edits are more difficult to automatically evaluate (56\% for motion-only edits). 
We hypothesize that this may be due to both models having only elementary understanding of motion, and being more biased towards spatial features of videos. 

For further comparisons of our method and baselines, we use the two metrics with highest observed correlations, $\text{M}_\text{dir}$ and $\text{M}_\text{geo}$. Results are shown in \Cref{table:results_automatic}, where our method outperforms all baseline methods. \Cref{table:results_metric_split} in the Supplementary shows a more detailed breakdown of evaluation results by edit type. However, in general, we note that due to the relatively low correlation between automatic metrics and human ratings, human judgements are more reliable in these evaluations. 

\vspace{0.5cm}

\begin{minipage}{\textwidth}
\centering
    \begin{minipage}[b]{0.5\hsize}\centering
            \includegraphics[width=\textwidth]{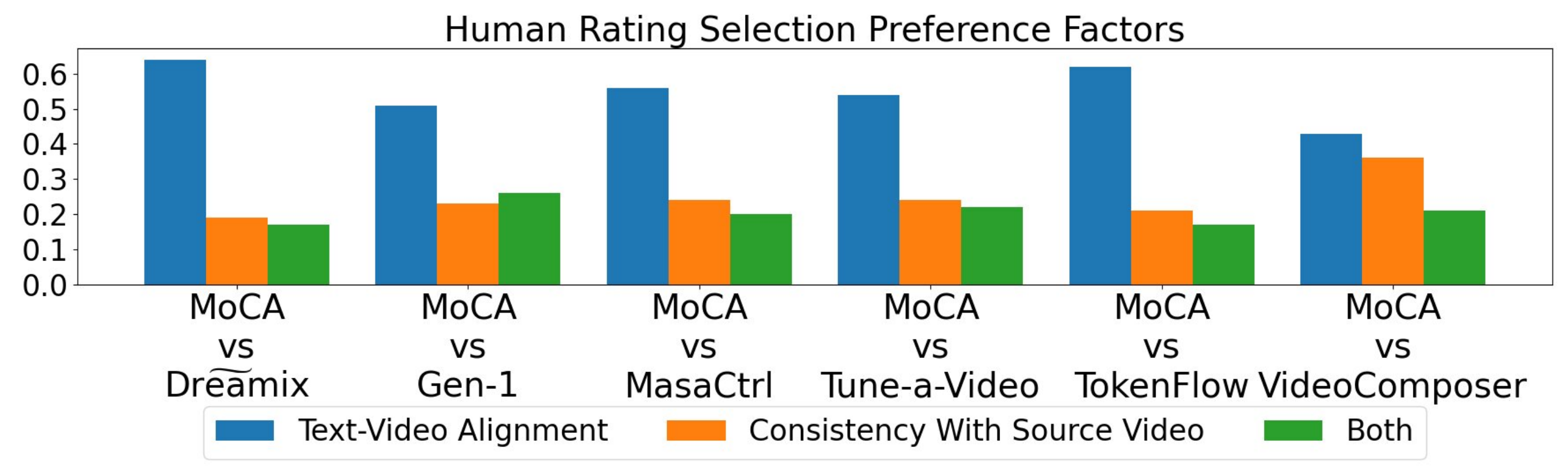}
    \captionof{figure}{\textbf{Percentage of each reason selected when human evaluators prefer \ours edits to each of the baselines.} The reasons for picking one model over another on each video edit could be either its better alignment with the edit prompt, higher consistency with the source video, or both. Generally, human raters preferred our method in terms of better alignment with the desired edit prompt.}
    \label{fig:human_factor_ratings}

    \vspace{0.7cm}

        \begin{tabular}{@{}lcc@{}}
        \firsthline
        \toprule
        Method                       & $\mathbf{M_{dir}}(\uparrow)$ & $\mathbf{M_{geo}}(\uparrow)$ \\ \midrule
        \textbf{\ours}          & \textbf{0.145}           & \textbf{0.301}           \\
        \textbf{\dreamix~\cite{dreamix}}       & 0.107                    & 0.252                    \\
        \textbf{Gen-1~\cite{gen1}}         & 0.111                    & 0.254                    \\
        \textbf{MasaCtrl~\cite{masactrl}}      & 0.090                    & 0.231                    \\
        \textbf{Tune-a-Video~\cite{tav}}  & 0.116                    & 0.265                    \\
        \textbf{TokenFlow~\cite{tokenflow}}     & 0.098                    & 0.235                    \\
        \textbf{VideoComposer~\cite{videocomposer}} & 0.128                    & 0.278                    \\ \bottomrule
        \end{tabular}
        \captionof{table}{\textbf{Automatic scores evaluating the editing quality of each model.} We compute the VideoCILP-based $\text{M}_\text{dir}$ and $\text{M}_\text{geo}$ scores as the CLIP text-video directional and geometric similarity scores, respectively, averaged across all edit tasks for each baseline. Our method shows higher editing capabilities compared to the baseline methods.}
        \label{table:results_automatic}  
        \vspace{0.5cm}

    \end{minipage}
            \hfill
    \begin{minipage}[b]{0.45\hsize}\centering
        \includegraphics[width=\textwidth]{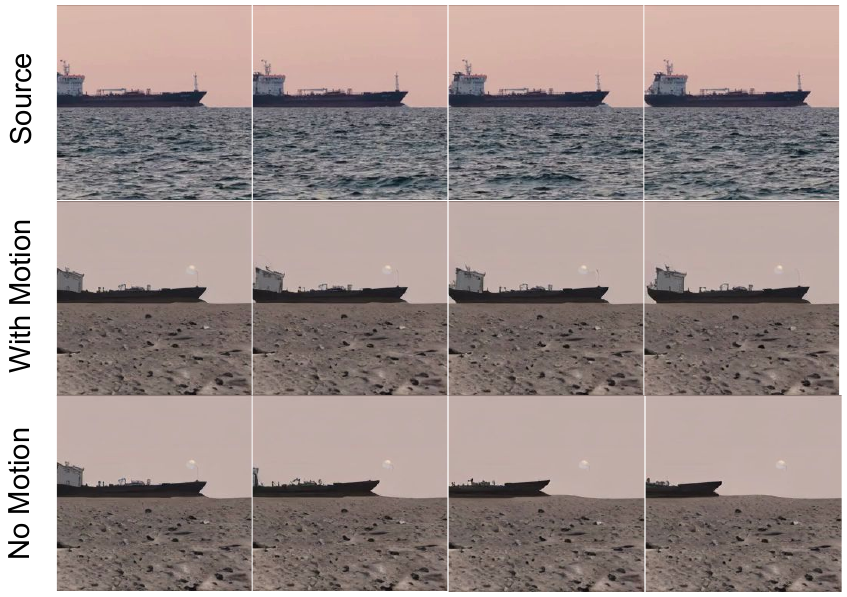}
        \captionof{figure}{\textbf{\ours edits for ``A boat sailing on the \textbf{moon}" with and without motion conditioning.} Using motion conditioning allows the model to more faithfully follow the boat's movement in the original source video. Without motion conditioning, the model tends to generate more random movement directions, such as moving backwards.}
        \label{fig:motion_comparison}
    \vspace{0.7cm}
        \begin{tabular}{@{}lp{0.5cm}p{0.7cm}p{0.9   cm}p{0.8cm}p{0.5cm}@{}}
        \firsthline
        \toprule
      & \textbf{Style} & \textbf{Object} & \textbf{\begin{tabular}[c]{@{}c@{}}Back-\\ ground\end{tabular}} & \textbf{\begin{tabular}[c]{@{}c@{}}Multi \\ Spatial\end{tabular}} & \textbf{Total} \\ \midrule
        $s_M=0$ & 57\%           & 60\%            & 57\%                                                            & 57\%                                                              & 58\%           \\ \bottomrule
        \end{tabular}
        \captionof{table}{\textbf{Ablation study on the motion conditioning in \ours} comparing the video edits conditioned on the motion of the source video against those without any motion conditioning. Human raters show a preference to our model with motion conditioning.}
        \label{table:ablation_no_motion}   

    \end{minipage}%
\end{minipage}

\subsection{Effect of Motion Conditioning} 
Lastly, we perform an ablation study on the motion conditioning introduced in our method. \Cref{table:ablation_no_motion} shows a comparison between our method with and without the motion conditioning support. Both models are trained on the same data for the same amount of iterations. We only perform evaluations on a subset of edit types (Style, Object, Background, Multi-Spatial) as our method does not use motion conditioning for the other motion-based edits ($s_M = 0$). For spatial edits, we find we are able to better preserve the original motion of the entities through motion conditioning. \Cref{fig:motion_comparison} Shows an example of when motion conditioning is beneficial in our model.

\section{Discussion}
\label{sec:discussion}
We introduced \ours, a method that decomposes the video editing problem into spatial and temporal components. Spatial edits are applied to the first frame of a source video, and then extrapolated using a motion-conditioned image animation model to preserve the motion of the original video. In addition, we allow motion editing by removing the motion conditioning and letting the animation model generate new frames according to the motion described in the edit prompt. We demonstrate that this simple method is a strong baseline outperforming existing methods on video editing. In addition, we introduce a new curated subset of video edits focused on motion editing, as well as a comprehensive analysis and benchmarking across a wide range of other video edits. By providing this comprehensive framework, we aim to facilitate the assessment of advancements and abilities of video editing techniques in subsequent research. We identify several limitations as directions for future work.

\begin{itemize}
    \item Analysis in \Cref{sec:automatic_evals} showed that all existing evaluation metrics for video editing are rather lacking in their alignment with human judgement. As such, there remains room for developing more accurate evaluation metrics for video editing, as human evaluations can be time consuming or expensive when using crowd sourced workers. In addition, a strong automatic metric may be useful for automatic selection of desired edits for hyperparameter searching or when comparing results from different random seeds.
    \item Due to our reliance on video extrapolation as means for video editing, our method has less fidelity when preserving any aspects of source videos that is introduced after the first frame, such as longer videos, or videos with more camera motion. Further work may involve incorporating other conditioning schemes that aim to preserve these parts of source videos, similar to our proposal of augmenting a video extrapolation model with motion conditioning to preserve motion changes.
\end{itemize}

\clearpage
\bibliographystyle{plainnat}
\bibliography{paper}

\clearpage
\beginappendix

\section{Qualitative Examples}

\begin{figure}[H]
    \centering
    \includegraphics[width=\textwidth]{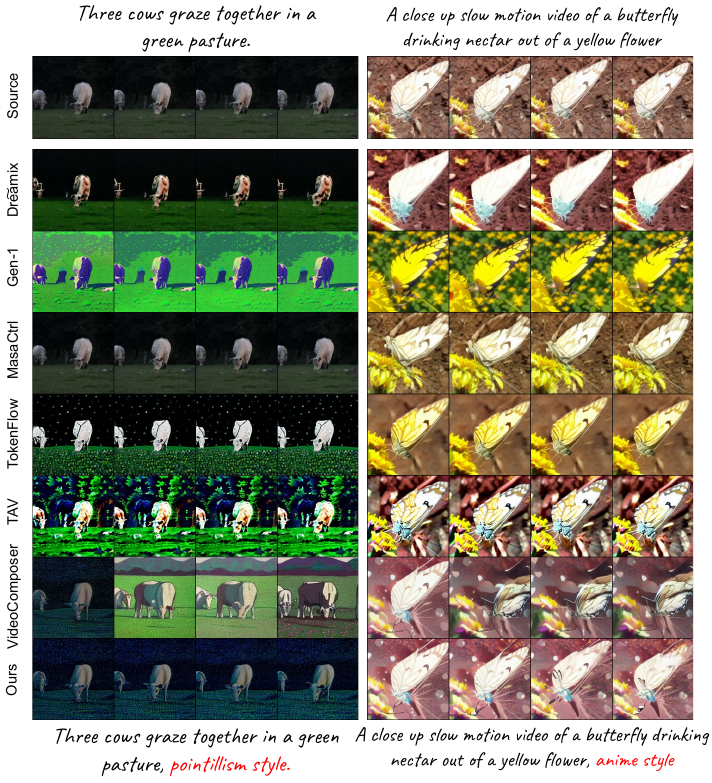}
    \caption{Comparisons for style video edit prompts}
\end{figure}
\begin{figure}[H]
    \centering
    \includegraphics[width=\textwidth]{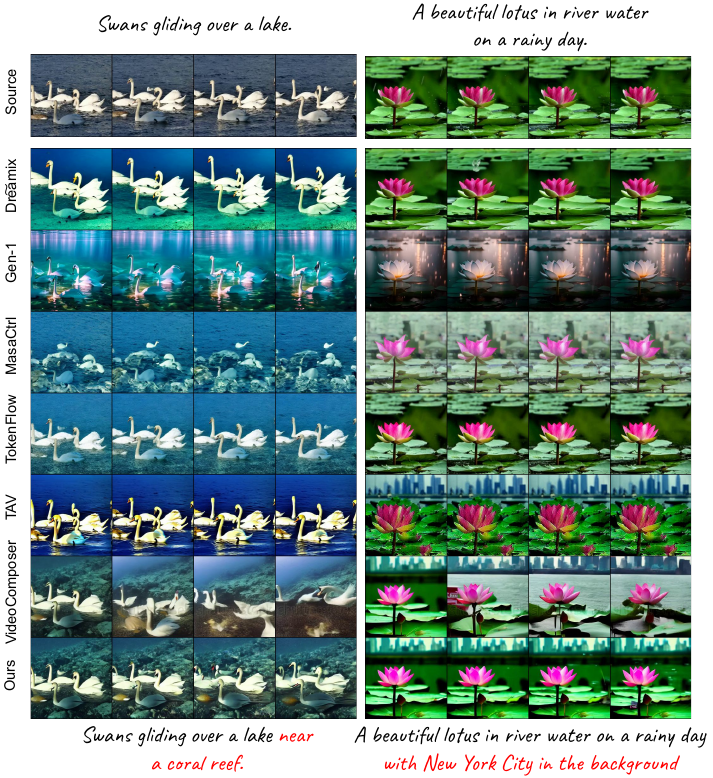}
    \caption{Comparisons for background video edit prompts}
\end{figure}
\begin{figure}[H]
    \centering
    \includegraphics[width=\textwidth]{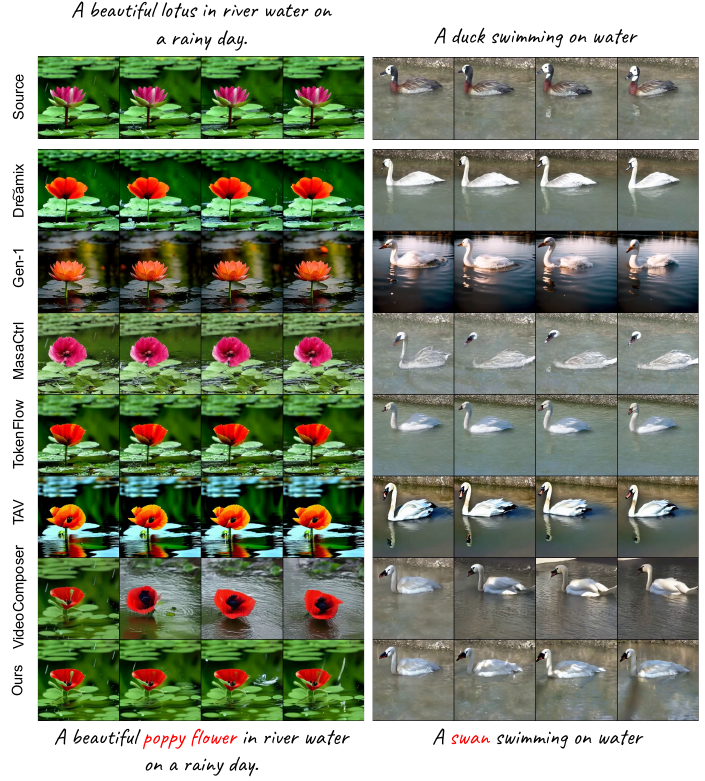}
    \caption{Comparisons for object video edit prompts}
\end{figure}
\begin{figure}[H]
    \centering
    \includegraphics[width=\textwidth]{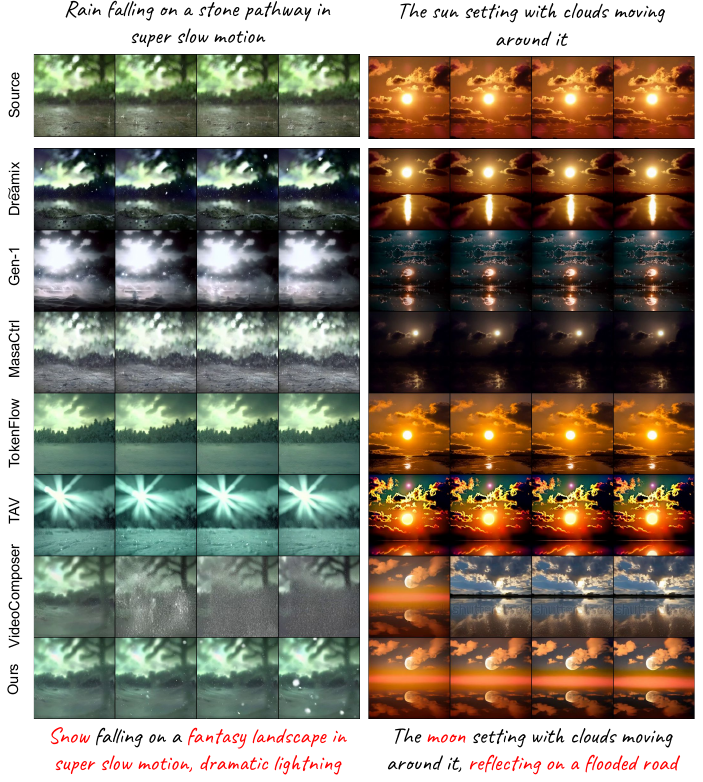}
    \caption{Comparisons for multi-spatial video edit prompts}
\end{figure}
\begin{figure}[H]
    \centering
    \includegraphics[width=\textwidth]{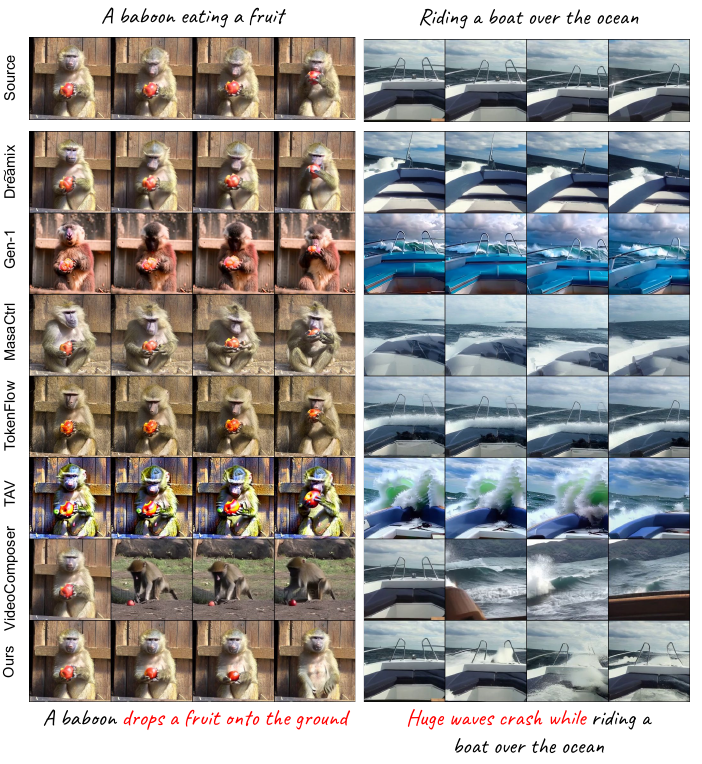}
    \caption{Comparisons for motion video edit prompts}
\end{figure}
\begin{figure}[H]
    \centering
    \includegraphics[width=\textwidth]{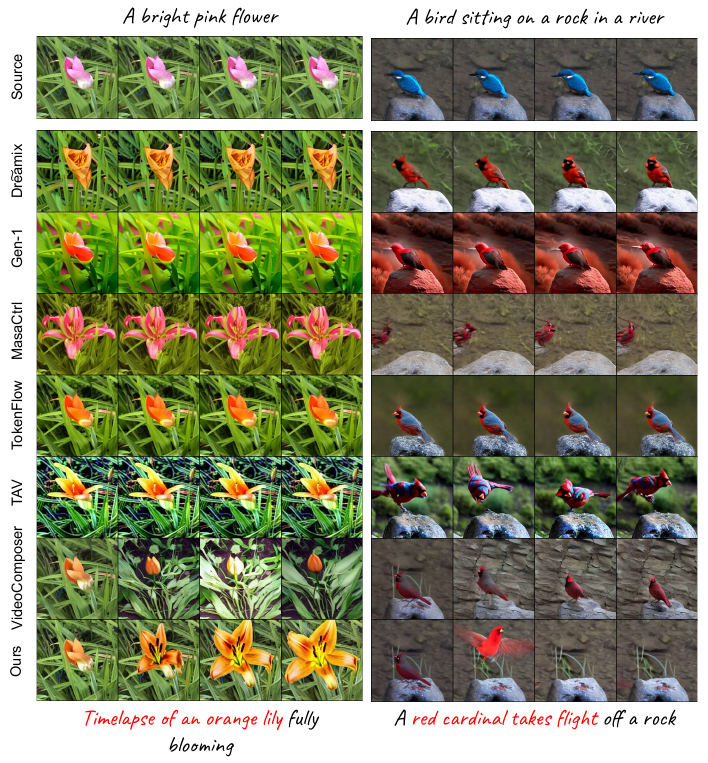}
    \caption{Comparisons for multi-motion video edit prompts}
\end{figure}
\newpage
\section{Automatic Evaluation Results}
We provide a more detailed comparison on different video editing methods using the VideoCLIP-based $\text{M}_{\text{geo}}$ metric in \Cref{table:results_metric_split}. The results confirm the superiority of \ours to other methods for all edit tasks. Additionally, we analyze the Spearman correlation~\cite{spearman} between automatic metrics introduced in \Cref{sec:automatic_evals} and human judgements. The results in \Cref{table:results_automatic_correlation} suggest the VideoCLIP based $\text{M}_{\text{geo}}$ and $\text{M}_{\text{dir}}$ metrics as the most reliable automatic metric in evaluating the performance of video editing models.

\begin{table}[h]
\centering
\begin{tabular}{@{}clccccccc@{}}
\toprule
\multicolumn{1}{l}{}        & \multicolumn{1}{c}{} & \textbf{Style} & \textbf{Background} & \textbf{Object} & \multicolumn{1}{l}{\textbf{Motion}} & \multicolumn{1}{l}{\textbf{Multi-Spatial}} & \multicolumn{1}{l}{\textbf{Multi-Motion}} & \multicolumn{1}{l}{\textbf{Total}} \\ \midrule
\multirow{4}{*}{ImageCLIP} & $\text{M}_\text{sim}$            & -0.095         & -0.102              & -0.042          & 0.174                               & -0.076                                     & -0.025                                    & -0.006                             \\
                            & $\text{M}_\text{dir}$      & 0.238          & 0.290               & 0.161           & 0.035                               & 0.219                                      & 0.141                                     & 0.150                              \\
                            & $\text{M}_\text{geo}$      & 0.240          & 0.288               & 0.156           & 0.047                               & 0.209                                      & 0.137                                     & 0.152                              \\ \midrule
\multirow{4}{*}{VideoCLIP} & $\text{M}_\text{sim}$            & -0.080         & -0.128              & 0.010           & 0.323                               & -0.080                                     & 0.006                                     & 0.036                              \\
                            & $\text{M}_\text{dir}$      & 0.290          & 0.328               & 0.183           & 0.049                               & 0.202                                      & 0.189                                     & 0.189                              \\
                            & $\text{M}_\text{geo}$      & 0.300          & 0.304               & 0.189           & 0.140                               & 0.201                                      & 0.205                                     & 0.203                              \\ \bottomrule
\end{tabular}
\caption{\textbf{\cite{spearman} correlation measuring the alignment between human judgements for editing quality and various CLIP-based metrics.} Among all six metrics in the table, $\text{M}_\text{geo}$ and $\text{M}_\text{dir}$ scores computed with a VideoCLIP model show the highest correlation with human ratings. Trends are similar to classification results shown in \Cref{table:metric_classification}, with higher correlation in spatial edits, and lower for motion-based edits.}
\label{table:results_automatic_correlation}
\end{table}

\begin{table}[h]
\centering
\begin{tabular}{@{}lcccccc@{}}
\toprule
Method & Style          & Background     & Object         & Motion         & Multi-Spatial  & Multi-Motion   \\ \midrule
\ours                                & \textbf{0.331} & \textbf{0.375} & \textbf{0.370} & 0.185 & \textbf{0.349} & \textbf{0.334} \\
{\dreamix}                             & 0.2223         & 0.304          & 0.356          & 0.141          & 0.290          & 0.321          \\
Gen-1                               & 0.254          & 0.317          & 0.295          & 0.146          & 0.309          & 0.209          \\
MasaCtrl                            & 0.225          & 0.253          & 0.295          & 0.154          & 0.270          & 0.283          \\
Tune-a-Video                        & 0.223          & 0.261          & 0.346          & 0.164          & 0.303          & 0.273          \\
TokenFlow                           & 0.206          & 0.239          & 0.314          & 0.0963         & 0.301          & 0.226          \\
VideoComposer                       & 0.259          & 0.328          & 0.326          & \textbf{0.187} & 0.301          & 0.202          \\ \bottomrule
\end{tabular}
\caption{\textbf{$\text{M}_\text{geo}$ metric computed for each model based on VideoCLIP features, averaged over all edit examples per manipulation type.} This table presents a more detailed split of results shown in \Cref{table:results_automatic} by edit type.}
\label{table:results_metric_split}
\end{table}

\end{document}